\documentclass[useAMS,usenatbib]{mn2e}

\usepackage{graphicx}\usepackage{epsfig}\usepackage{epsf}
\usepackage{amsmath}\usepackage{amssymb}\usepackage{stfloats}
\title[The CO torus of Eta Car]{A disrupted molecular torus around Eta
  Carinae as seen in $^{12}$CO with ALMA}

\author[Smith]{Nathan Smith$^{1}$\thanks{E-mail:
    nathans@as.arizona.edu}, Adam Ginsburg$^{2,3}$, and John Bally$^4$
  \\ $^{1}$Steward Observatory, University of Arizona, 933 N. Cherry
  Ave., Tucson, AZ 85721, USA \\ $^2$Jansky Fellow of the National
  Radio Astronomy Observatory, Socorro, NM 87801 USA \\ $^3$European
  Southern Observatory, Karl-Schwarzschild-Straße 2, D-85748 Garching
  bei M\"{u}nchen, Germany \\ $^4$Center for Astrophysics and Space
  Astronomy, University of Colorado, Boulder, CO 80309, USA}

\begin{document}

\pagerange{\pageref{firstpage}--\pageref{lastpage}} \pubyear{2012}
\maketitle
\label{firstpage}

\begin{abstract}

  We present Atacama Large Millimeter Array (ALMA) observations of
  $^{12}$CO 2$-$1 emission from circumstellar material around the
  massive star $\eta$~Carinae.  These observations reveal new
  structural details about the cool equatorial torus located
  $\sim$4000 au from the star. The CO torus is not a complete
  azimuthal loop, but rather, is missing its near side, which appears
  to have been cleared away. The missing material matches the
  direction of apastron in the eccentric binary system, making it
  likely that $\eta$~Car's companion played an important role in
  disrupting portions of the torus soon after ejection.  Molecular gas
  seen in ALMA data aligns well with the cool dust around $\eta$~Car
  previously observed in mid-infrared (IR) maps, whereas hot dust
  resides at the inner surface of the molecular torus. The CO also
  coincides with the spatial and velocity structure of near-IR H$_2$
  emission.  Together, these suggest that the CO torus seen by ALMA is
  actually the pinched waist of the Homunculus polar lobes, which
  glows brightly because it is close to the star and warmer than the
  poles. The near side of the torus appears to be a blowout,
  associated with fragmented equatorial ejecta. We discuss
  implications for the origin of various features northwest of the
  star.  CO emission from the main torus implies a total gas mass in
  the range of 0.2 - 1 $M_{\odot}$ (possibly up to 5 $M_{\odot}$ or
  more, although with questionable assumptions).  Deeper observations
  are needed to constrain CO emission from the cool polar lobes.

\end{abstract}


\section{INTRODUCTION}

The Homunculus around the massive binary star $\eta$~Carinae is a
striking example of a bipolar circumstellar nebula with a tightly
pinched waist (see \citealt{smith12} for a recent overview).  It
serves as a prototypical object in ongoing attempts to understand the
shaping of bipolar nebulae, and their physical orgin from close binary
systems or rapidly rotating stars \citep{f99}.  It is perhaps most
recognizable from stunning visible-wavelength images with the {\it
  Hubble Space Telescope} ({\it HST}) that show the complex clumpy
structure of the dust shell \citep{morse98}, which unlike most ionized
planetary nebulae, is seen primarily in reflected light at visible
wavelengths.  However, many clues about its structure and physical
properties come from longer wavelengths where the nebula becomes
transparent.

The dust shell around the star absorbs much of the star's tremendous
luminosity and reradiates that energy in the infrared (IR), making the
Homunculus around $\eta$ Car one of the brightest extrasolar sources
in the sky at thermal IR wavelengths \citep{nw68,wn69}.  This made it
a favorite target for early ground-based mid-IR observations, with
repeated study spanning several decades as resolution and sensitivity
improved
\citep{rdg73,rob73,sut74,joy75,ait75,apru75,andr78,mit78,hyl79,mit83,chel83,bens85,hgg86,russ87,allen89,smetal95,s+98,pol99,morr99,pant00,hony01,smith02,smith03,ches05}.
Most of these showed an inner structure that was elongated
perpendicular to the bipolar axis, often interpreted as a
limb-brightened equatorial torus.  Many of these authors supposed that
such a dense equatorial torus might have been the agent that pinched
the waist of the nebula, as seen in numerical simulations
\citep{f+95,f+98,db98,lang99}.  The caps of the polar lobes are very
dense and massive \citep{smith03,smith06}, while the side walls and
equatorial parts of the Homunculus are relatively transparent
\citep{kd01,smith02,smith06}, so it would be difficult to make such a
model work unless there is a large mass reservoir of cold gas hidden
in dense equatorial regions.  While mid-IR emission traces warm dust
heated by direct radiation from the central star, shielded dense gas
could be cold and might be revealed by molecular emission at longer
wavelengths.

Compared to the long history of studying its IR emission, molecular
gas in the circumstellar nebula of $\eta$ Car was detected fairly
recently.  Early surveys of CO in the Carina Nebula did not detect CO
line emission from $\eta$ Carinae itelf \citep{cox95}.  The first
reported detection of molecules in the Homunculus arose via its
near-IR emission from H$_2$ in several lines around 2~$\mu$m
\citep{sd01,smith02}.  This H$_2$ emission traced very dense and very
thin walls of the polar lobes of the Homunculus
\citep{smith02,smith04,smith06,sf07}, but was not seen from the
equatorial ejecta except at the pinched waist of the polar lobes.
Absorption from H$_2$ and several diatomic molecules like CH and OH
was detected in UV spectroscopy of $\eta$ Car obtained with {\it HST},
where the radial velocity and narrow line width indicated that this
molecular gas was located in the thin walls of the south-east polar
lobe \citep{niel05,ver05}.  The first report of a polyatomic molecule
was ammonia via the NH$_3$ ($J$,$K$) = (3,3) inversion transition
\citep{sNH3}, although we note that Loinard (priv.\ comm.) suggests
that this line may actually be H recombination.  Several additional
molecules have since been detected via single-dish observations from
the inner regions of the Homunculus
\citep{loinard12,loinard16,morris17}, and have also been attributed to
the dusty torus seen in the IR.  CO emission has also been detected in
the circumstellar environments of other LBVs, including AG Carinae and
HR Carinae \citep{nota02,mcgregor88}.  The total molecular gas mass
around AG Car was estimated to be roughly 3 $M_{\odot}$, thought to
reside mostly in an equatorial belt \citep{nota02}.  In $\eta$~Car,
the H$_2$ emission is located mostly in the walls and caps of the two
large polar lobes, but the density structure of the molecular gas
located in equatorial regions --- and its possible association with the
IR torus --- has remained uncertain.

Here we present new Atacama Large Millimeter Array (ALMA) observations
in the 230.5 GHz (1.3 mm) $^{12}$CO 2$-$1 line obtained with an
angular resolution of roughly 1$\arcsec$.  These observations clearly
delineate the spatial relationship between the densest molecular gas
in the Homunculus as compared to the dusty IR torus and the thin walls
of the polar lobes.  We present the new observations in Section 2, and
make various comparisons to spatially resolved optical and IR data in
Section 3.

\begin{figure*}
\includegraphics[width=5.8in]{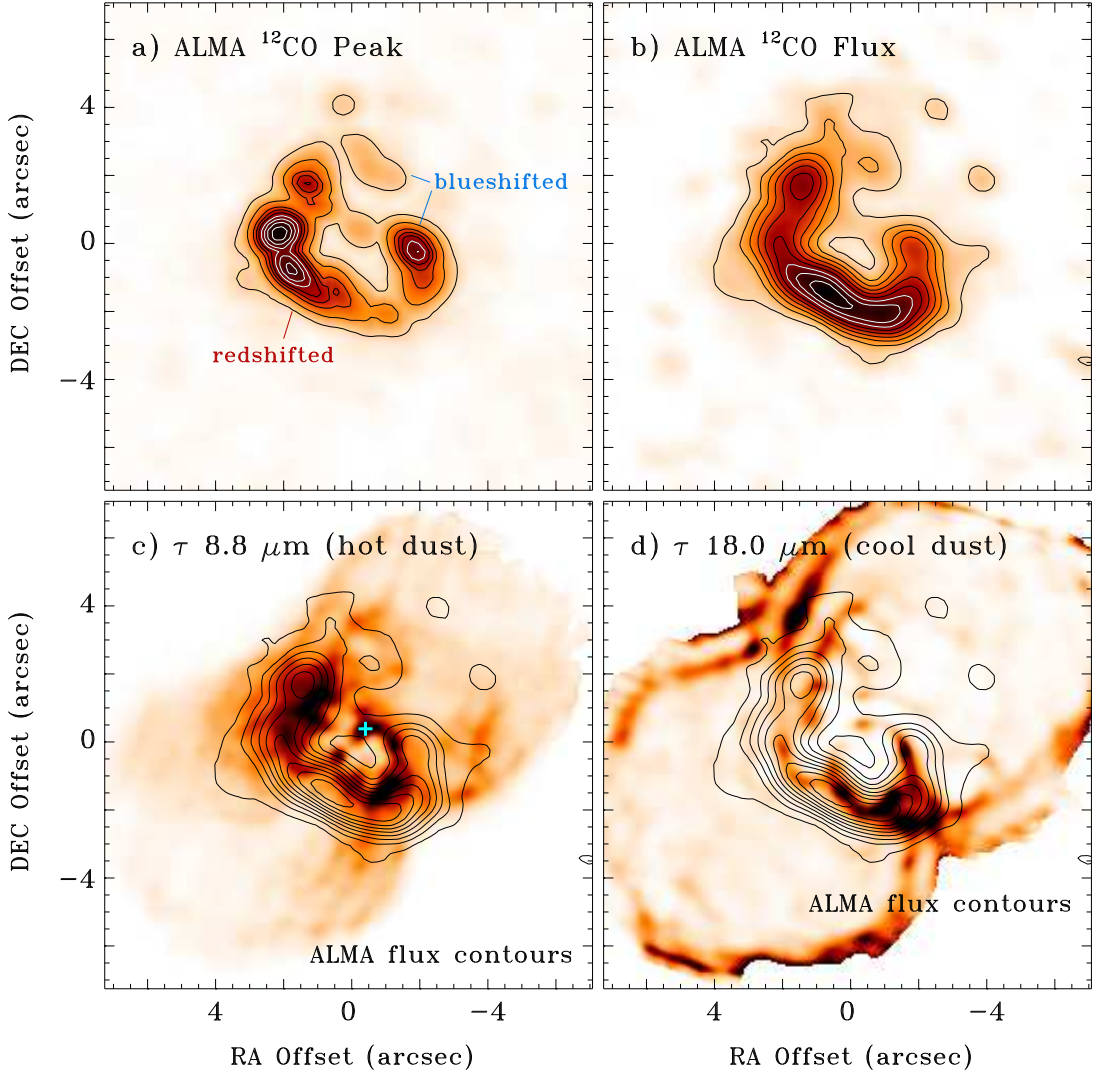}
\caption{The top two panels show maps of the peak emission (a) and
  zeroth moment or integrated line flux (b) from the ALMA $^{12}$CO
  2-1 data cube.  RA and DEC offsets are relative to the star, for
  which the registration uncertainty is about 0$\farcs$5. Countours in
  Panel (a) are at 10, 20, 30, 35, 40, 45, 50, 55, and 60 mJy.
  Contours in (b) are at 200, 400, 600, 800, 1000, 1200, 1400, 1600,
  and 1800 mJy km/s.  The Jy to K conversion factor is 18.1 Jy
  K$^{-1}$, using a 1$\farcs$28$\times$0$\farcs$99 beam at P.A. = 67
  deg at 230.5 GHz.  The contours from (b) are superposed on the two
  images in the bottom panels.  These images show the 8.8 $\mu$m
  optical depth (esentially the column density of hot 200-400 K dust)
  in Panel (c), and the 18 $\mu$m optical depth (essentially the
  column density of cooler 100-200 K dust) in Panel (d).  More details
  on the IR dust maps can be found in \citet{smith+02} and
  \citet{smith03}.  The blue ``+'' in panel (c) marks the hot NW dust
  condensation discussed in the text.  }
\label{fig:IR}
\end{figure*}

\section{OBSERVATIONS}

ALMA observations of $\eta$~Carinae in Band 6 (1.3 mm) were taken as
part of program 2013.1.00661.S.  Two scheduling blocks were executed
on 2015 January 29 and 2015 April 3. The observations covered 4
basebands: 215.7$-$217.6 GHz, 217.6$-$219.5 GHz, 229.6$-$231.5 GHz,
and 231.6$-$233.5 GHz. The velocity resolution achieved was 1.3 km
s$^{-1}$ (976 kHz).  Data reduction was performed using the CASA
pipeline version 4.2.2 r30986, and we used the QA2-produced
visibilities and images. The continuum emission was fit and
subtracted.  Here we concentrate on the 230.5 GHz line of $^{12}$CO
2$-$1.  For the $^{12}$CO 2$-$1 line, the RMS achieved was 7.2 mJy per
3 km s$^{-1}$ channel in a 1$\farcs$3$\times$0$\farcs$99 \ (P.A. =
67$^{\circ}$) beam.  Images were produced with Briggs weighting using
robust factor 0.5 and a cell size of 0$\farcs$2.

Figure~\ref{fig:IR} shows maps of the peak CO emission
(Fig.~\ref{fig:IR}a) and the integrated line flux or 0th moment of the
CO emission (Fig.~\ref{fig:IR}b). Panels (c) and (d) show the same
contours of line flux from Fig.~\ref{fig:IR}b, but superposed on
images of the mid-IR 8.8~$\mu$m and 18~$\mu$m optical depth maps,
which show the relatively hot and cool mid-IR dust column density,
respectively \citep[from][]{smith03}.  Figure~\ref{fig:H2} shows a
position velocity diagram of the CO emission, and compares it to
high-resolution near-IR spectra of near-IR H$_2$ emission
\citep{smith04,smith06}.  Figure~\ref{fig:HST} shows the same peak and
line flux contours (from Figs.~\ref{fig:IR}a and b) superposed on a
near-UV {\it HST} image of the Homunculus, for reference.

\section{RESULTS AND DISCUSSION}

\subsection{The CO and IR Torus}

Images of the peak $^{12}$CO emission and the integrated line flux
(Figures~\ref{fig:IR}a and b, respectively), show a clumpy ring-like
structure, with an appearance that is very reminiscent of the IR torus
structure in the core of the Homunculus, as discussed in the
Introduction.  While these two maps trace the same basic structure,
there are subtle differences due to the velocities over which the
emission is dispersed. The apparent ring seen on the sky is indeed
consistent with an expanding torus in the equatorial plane of the
Homunculus, because it is redshifted to the southeast (SE) and
blueshifted to the northwest (NW).  This gives an orientation
perpendicular to the expansion of the polar lobes, consistent with
expansion in the equator.  Correcting for inclination, the radial
expansion of this torus is 100 to 200 km s$^{-1}$ (see below).

This torus, however, is not a smooth and azimuthally complete ring
surrounding the star.  The brightness structure reveals a handful of
large condensations, with a pronounced wide gap in the ring or torus
on the blueshifted NW side.  In both the peak and integrated flux
maps, the torus exhibits a ``C'' shape, with the opening toward the
NW.  The integrated flux map gives the impression that the CO mass is
concentrated on the redshifted SE side of the torus, especially
apparent in Figure~\ref{fig:IR}b.  The clumpy and broken structure of
this torus is remarkably similar, qualitatively, to the clumpy
``C''-shaped density distribution in the ionized ring nebula around
the massive contact binary RY Scuti \citep{smith02b}. The strong
departure from azimuthal symmetry in the CO torus around $\eta$~Car
may be extremely important for understanding the observed structure
and origin of the Homunculus, as discussed more below.

The initial impression that the CO torus is reminiscent of the IR
torus discussed in decades past is accurate.  In fact, they have
roughly the same size and structure.  Figures~\ref{fig:IR}c and d show
the CO line flux contours superposed on mid-IR optical depth maps of
the relatively warm (200-250 K) and cool (140 K) dust
\citep{smith+02,smith03}.  The clumpy IR torus shows good spatial
correspondence with the CO torus in terms of spatial structure; they
should not match in the brightness distribution, since the CO map is
flux and the IR maps are optical depth.  (Note that maps of the IR
optical depth essentially trace the dust mass column at different
temperatures, rather than the brightness distribution.  The central
core region of the Homunculus is brighter at all IR wavelengths,
because that is where the dust is hottest.)  

Both the spatially resolved dust temperature and IR emission-lines
indicated a stratified double-shell structure in the Homunculus, with
an inner zone of hotter dust and low-ionization atomic emission lines,
and with a thin outer zone of cooler dust and H$_2$ emission
\citep{smith02,smith06,smith03}.  The sptially resolved CO emission in
our ALMA data arises from the outer of these two zones.  Compared to
the hot dust (traced by 8.8 $\mu$m optical depth), CO peaks at
somewhat larger separation from the star, about 1{\arcsec} outside the
8.8 $\mu$m optical depth peak (Figure~\ref{fig:IR}c).  However, the CO
structures match well with the locations of several peaks in the 18
$\mu$m optical depth map (cooler dust; Figure~\ref{fig:IR}d).  This is
expected if the CO resides primarily in the walls of the polar lobes
of the Homunculus, like H$_2$, because the thin walls are known to
have a stratified temperature and density structure
\citep{smith03,smith06,sf07}.  The spatial offset between the
locations of the warmest dust and the CO is only about
0$\farcs$5$-$1$\arcsec$, comparable to the thickness of the walls of
the polar lobes themselves \citep{smith06}.  The hotter dust seen in
the 8.8 $\mu$m emission traces a thin skin on the inside walls of the
hollow polar lobes exposed to direct starlight, whereas the cooler
dust and CO is shielded by the dense walls of the polar lobes and
resides at slightly larger distances from the star.

Located within the gap in the torus (the opening of the ``C'') to the
NW of the star, there is a feature that shows a peak in the column of
hot dust (in Fig.~\ref{fig:IR}c), but that is much weaker in the cool
dust and CO maps.  This is a particularly noteworthy location, and is
marked with a blue ``+'' in Figure~\ref{fig:IR}c.  It traces a complex
of dust knots that marks the peak brightness in mid-IR images of the
Homunculus, because there is a complex of dusty knots close to the
star where the dust is heated to high temperatures, and where the
inside edges of these dust clumps are ionized by UV radiation.  Some
of the brightest of these ionized knots are known as the ``Weigelt
knots'' \citep{weigelt86,weigelt95}, which have received much
observational attention, although images show several other knots in
the vicinity \citep{smith04a,ches05,gull16}.  This dust feature is
found in what is otherwise a gap in the CO and cool dust torus --- yet
this clump appears to have remained.  Based on its structure,
different temperature and column density, and kinematic evidence, we
suspect that its location as part of the hot dust torus may be
misleading, and that it may actually have a different origin, as
discussed later.

Excluding this hot dust immediately NW of the star, the wide gap on
the NW side of the torus is especially apparent in CO emission and the
cool dust optical depth map in Figure~\ref{fig:IR}d.  These maps
create the impression that this portion of the torus in the
blueshifted side has suffered a blow-out, as if someting has disrupted
the torus in this direction.  Interestingly, this is a special
direction in the $\eta$ Car system, as discussed later.  First we
discuss the kinematic structure of the CO emission detected by ALMA.

\begin{figure}
\includegraphics[width=3.0in]{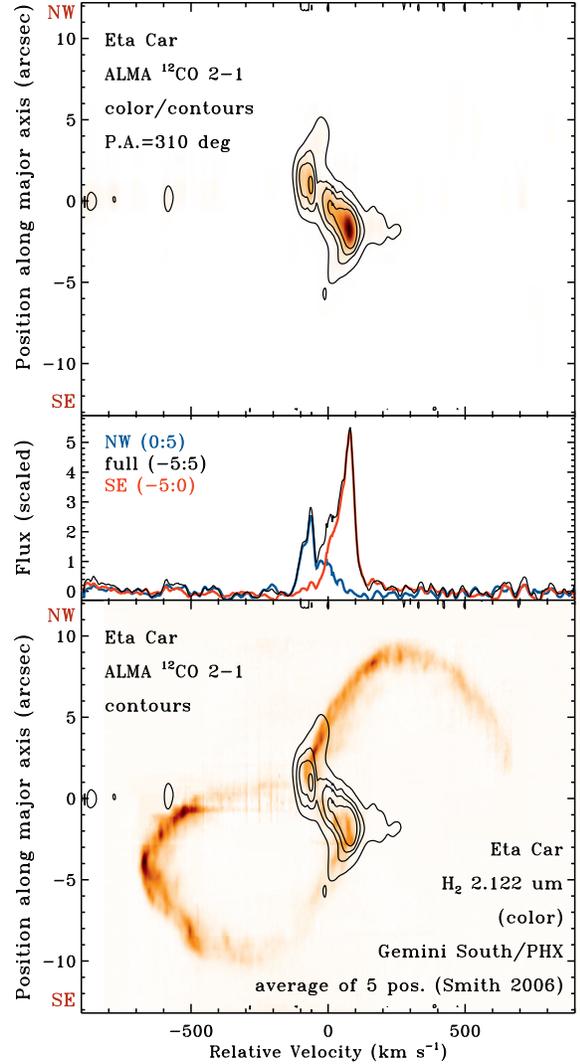}
\caption{{\it Top:} Position-velocity diagram of the ALMA $^{12}$CO
  2-1 data with the position running along the major axis of the
  Homunculus at P.A. = $-$50$^{\circ}$, integrated over a spatial
  width of 5{\arcsec} (across the width of the ``torus''). {\it
    Middle:} Integrated spectrum of the CO emission.  The black line
  is emission from the full torus integrated over $-$5 to $+$5 arcsec
  along the position axis in the top panel, the blue line is the
  blueshifted NW side of the torus integrated over 0 to $+$5 arcsec,
  and the red line is the redshifted SE part of the torus integrated
  over $-$5 to 0 arcsec.  {\it Bottom:} Same contours from the top
  panel, but superposed on the near-IR H$_2$ emission from the
  Homunculus, taken with the Phoenix spectrograph on Gemini South.
  This is the average of several slits oriented along the same
  position angle, from \citet{smith06}.}
\label{fig:H2}
\end{figure}

\subsection{Near-IR H$_2$ Emission and Kinematic Structure}

In the previous section, a comparison of the new CO emission to the
previously known dusty IR torus gave the impression that these two are
closely related.  The structure in images suggested that --- like the
IR torus -- the CO emission traces material in the thin walls of the
Homunculus, which is brightest in the pinched waist near the equator.

This interpretation is supported by the kinematics.
Figure~\ref{fig:H2} (top) shows a position-velocity diagram with the
positional coordinate along the major axis of the Homunculus, and with
the flux integrated in the direction perpendicular to the polar axis.
This indicates that the CO emission has a flattened distribution,
elongated perpendicular to the polar axis, in an expanding equatorial
torus.

Figure~\ref{fig:H2} (bottom) compares the kinematics of CO seen by
ALMA to the kinematics of molecular hydrogen, as seen in
high-resolution spectra of the Homunculus in the 2.122 $\mu$m H$_2$
line \citep{smith04,smith06}.  On the redshifted side of the CO torus
(and toward negative positional offsets in Figure~\ref{fig:IR}), where
the CO emission is brightest and where the CO torus traces the IR
torus (see above), the kinematic structure of the CO emission overlaps
perfectly with the near-IR $H_2$ emission.  This confirms the
conclusion from comparing the CO and IR maps (see above), that the CO
emission seen by ALMA is not a more extended disk structure separate
from the Homunculus, but is in fact just the molecular gas in the thin
walls of the polar lobes that are pinched at the equator.  The spatial
coincidence of the spatially resolved CO emission with both the cool
dust and H$_2$ emission from the outer shell of the Homunculus walls
indicates that the CO is a good tracer of the cooler, dense, shielded
zones in the Homunculus walls where most of the mass resides
\citep{sf07}.  The CO is not limited to a thin skin on the inner
surface of these walls of the Homunculus (where the hotter dust
resides), and so it is likely that the CO 2-1 emission is tracing the
regions with most of the mass.

The redshifted peak of the torus has a radial velocity of about $+$81
km s$^{-1}$, which corresponds to an expansion speed away from the
star (correcting for an inclination of $i$=41$^{\circ}$;
\citealt{smith06}) of about 123 km s$^{-1}$.  This is the same as the
H$_2$ emission, as noted above.  \citet{morris17} noted a similar
velocity range for far-IR transitions of CO, which suggests again that
both the H$_2$ and CO emission lines are probably optically thin and
are tracing the same gas.  This makes it unlikely that a huge
reservoir of cold, shielded mass would be undetected in our ALMA
data. The projected radius of the torus is $\sim$1.9 arcsec, or a
radial distance from the star of 4400 AU. This translates to a
kinematic age of around 170 ($\pm$15) yr, indicating that the CO torus
was ejected during the Great Eruption along with the rest of the
Homunculus \citep{morse01,smith17}.  The CO emission is not tracing an
older, pre-existing disk as advocated by \citet{morr99}.  The reason
that the CO appears only as a torus in the flux maps is probably just
because this gas is the closest to the star, and so is the warmest and
brightest of the CO emission.  We suspect that deeper CO observations
are needed to detect CO at larger Doppler shifts in the polar regions
of the lobes, where H$_2$ is seen.

\begin{figure*}
 \includegraphics[width=5.8in]{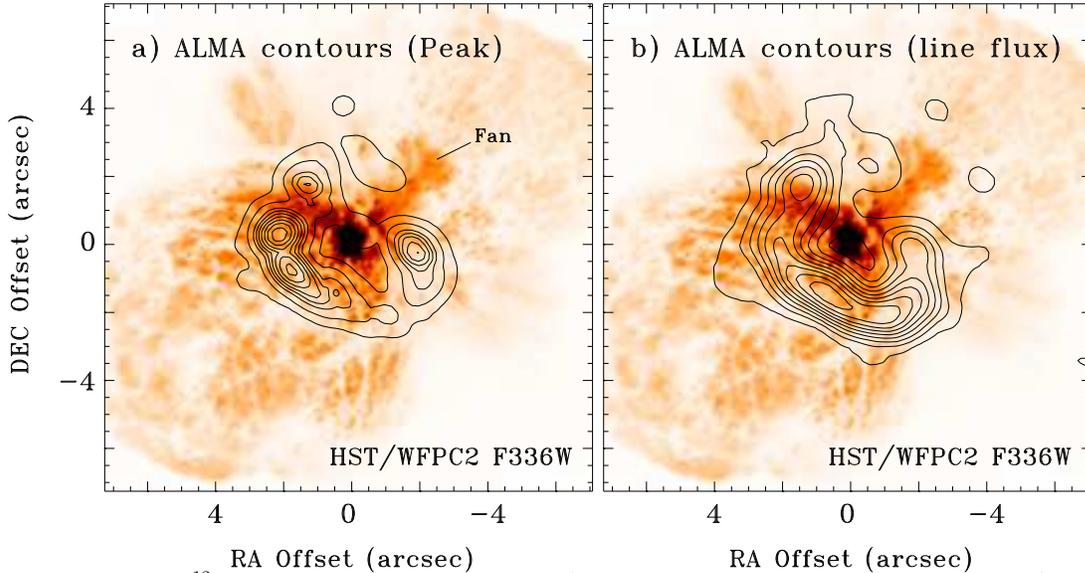}
\caption{These are the same $^{12}$CO contours from
  Figure~\ref{fig:IR}a and b (peak emission and zeroth moment,
  respectively), but superposed on the {\it HST}/WFPC2 F336W image from
  \citet{morse98}.}
\label{fig:HST}
\end{figure*}

On the blueshifted side of the CO emission distribution, which is
fainter (and located toward positive spatial offsets), most of the CO
also overlaps with the near-IR H$_2$ emission from the NW polar lobe
walls in Figure~\ref{fig:H2}.  There are, however, some fainter CO
clumps on the NW/blueshifted side of the torus that are farther from
the star, fainter, and more ragged; although these don't show up in
the integrated position-velocity diagram in Figure~\ref{fig:H2}, they
can be seen in Figure~\ref{fig:IR}.  CO structures appear to be less
dense and more wispy to the NW direction, as compared to the rest of
the torus.  Altogether, the IR and CO images give the impression that
the NW side of the torus has been disrupted or blown out, with thin
clumps pushed to larger distances from the star as something broke
through the torus.  This may be related to many of the interesting and
peculiar aspects of the equatorial ejecta to the NW of the star that
have been noted by many authors, as discussed in the next section.

\begin{figure*}
\includegraphics[width=5.8in]{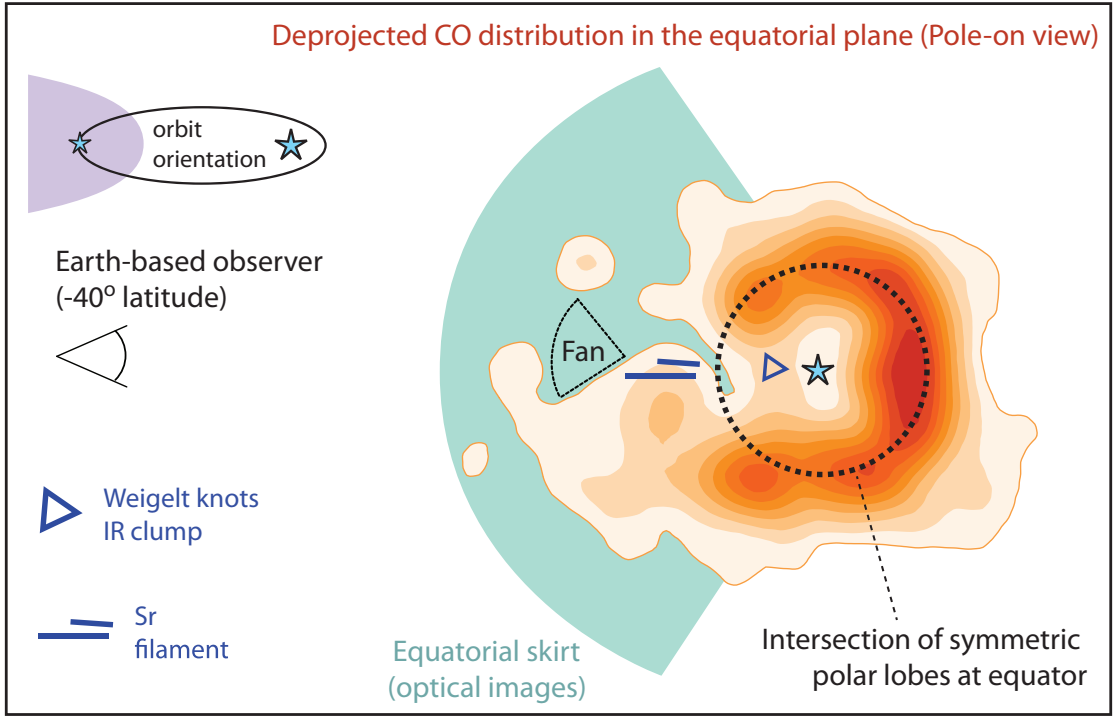}
\caption{Sketch of the equatorial geometry.  The red/orange contours
  correspond to the ALMA $^{12}$CO flux in an image, deprojected for
  an inclination angle of $i$$\simeq$40$^{\circ}$, corresponding to
  the tilt of the equatorial plane of the Homunculus from the plane of
  the sky \citep{smith06}.  This is an approximation of a ``pole-on''
  view of the CO as seen by an observer in the SE polar lobe.  The
  location of an Earth-based observer (who would be 40$^{\circ}$ out
  of this image plane) is at left.  Other features are drawn to
  represent the extent of the visible equatorial skirt (green pie
  wedge), the location of the Strontium filament (blue hash marks) and
  the complex of dusty knots that includes the Weigelt knots (blue
  triangle). The location of the ``Fan'' is also noted.  The
  approximate orientation of the central binary system's orbit is
  shown at the upper left for reference (with the purple zone
  representing the colliding-wind shock cone opening), where the
  direction of the secondary at apastron is to the left.}
\label{fig:sketch}
\end{figure*}

\subsection{The ``blowout'' on the near side of the torus, and its
  relation to the equatorial ejecta}

In the two previous subsections, we noted how the observed CO
structure in combination with other data gives the impression that the
dense equatorial material around $\eta$ Car has two key aspects.  The
first main component is an incomplete expanding torus with a
``C''-shaped density distribution, where the NW/blueshifted side of
the expanding torus is missing.  This torus corresponds to the thin,
cold walls of the Homunculus that are pinched at the waist, where
their proximity to the star makes them much brighter than the rest of
the nebula.  The second feature is a relative paucity of CO emission
on the blueshifted side of the equator corresponding to the gap in the
``C''-shaped density distribution, which might be a ``blowout'' where
molecular gas to the NW of the star has been pushed out or diverted,
and takes on a much more ragged or whispy distribution.

Figure~\ref{fig:HST} shows the CO contours superposed on an
optical/near-UV image from {\it HST} \citep{morse98}, showing the
reflection nebula and especially the equatorial ejecta.  While we of
course cannot see the far side of the equatorial region at visible
wavelengths because of obscuration, it is clear that on the near side,
the faint CO features are pushed to larger radii than the rest of the
CO/IR torus, concident with the region of the peculiar equatorial
ejecta.  In fact, the CO emission a few arcseconds to the N and E of
star corresponds to dark patches in the equatorial ejecta, which have
been inferred to be regions of higher visual-wavelength extinction in
the equatorial skirt based on multiwavelength optical/IR imaging
\citep{s+98,smith99,smith+02,smith03}.

The gap to the NW may hold vital clues and may have important
implications for understanding the time sequence of mass ejection that
produced the Homunculus, because the NW direction is special.  In the
central binary system, the NW corresponds to the direction of apastron
of the very eccentric orbit \citep{madura12,gull09}, where $\eta$
Car's hot companion spends most of its time during the 5.5 yr cycle
(see Figure~\ref{fig:sketch}).  The opening of the colliding-wind
shock cone therefore points in this direction most of the time.  There
are several notable nebular features (see Fig.~\ref{fig:sketch}) that
are all associated with this direction in the equatorial plane of the
Homunculus:

1) {\it The Weigelt knots and NW dust complex.}  This is a series of
dense condensations located 0$\farcs$1$-$0$\farcs$5 from the central
star that were resolved by early speckle interferometry techniques
\citep{weigelt86,weigelt95} and in high-resolution IR images
\citep{ches05}. These nebular knots have very strong narrow atomic
emission lines throughout the UV, optical, and near-IR spectrum
\citep{davidson95,davidson97,smith02,smith04a,hartman05}.  Proper
motions show that the Weigelt knots and some of the other features are
younger than the polar lobes \citep{dorland04,smith04a}, having been
ejected in the 1890 eruption or later.

2) {\it The bright ``Fan'' to the NW of the star}.  This feature
appears to squirt out through the gap in the torus (see
Fig.~\ref{fig:HST}a).  In fact, it has been noted previously that the
Fan is an optical illusion \citep{s+98,smith99,kd01,smith03,smith12}.
Rather than a density structure, it is bright because it is a patch of
lower extinction (i.e. a hole, rather than a dense clump in the
equator) allowing scattered light from the NW lobe to pass through the
equatorial plane.  This is consistent with the paucity of CO emission
at this location.

3) {\it Several thin filaments at the base of the Fan; the so-called
  ``strontium filament'' and others.}  These are a collection of
several thin filaments located 0$\farcs$5 to 2$\farcs$5 NW of the
star.  They show peculiar low-excitation, with strong emission from
unusual forbidden lines including [Sr~{\sc ii}] and many other
low-ionization metal lines \citep{hartman04,bautista06,bautista09}.
Radial velocies show peculiar kinematics that don't match simple
expectations for the equator of the Homunculus, either corresponding
to a range of ejection dates before and after the Great Eruption, or
instead, a range of latitudes out of the equator
\citep{zethson99,kd01,smith02}.

4) {\it Extended equatorial emission features in He~I and other
  lines.}  Emission from hot equatorial gas that may have multiple
ages (or multiple latitudes offset from the equator) are seen in lines
such as He~{\sc i} $\lambda$10830 in front of the NW polar lobe
\citep{smith02,smith08,teodoro08}.  Curiously, these features are not
seen in H~{\sc i} recombination lines or [Fe~{\sc ii}] lines like most
of the other extended ejecta outside the Homunculus.  This emission
appears to be projected outward from the gap in the CO torus.

5) {\it The equatorial skirt is one-sided.}  The collection of dusty
clumps, streaks, bright spots, and streamers that collectively make up
the ``equatorial skirt'' are prominent in near-UV and
visual-wavelength images in scattered light (Fig.~\ref{fig:HST}).
They seem to fade and then disappear as we move to longer wavelengths
in the near-IR and mid-IR, however, suggesting that they are mostly
scattered light illumination effects (i.e. searchlight beams escaping
the core) rather than dense structures in a disk
\citep{smith99,smith02,smith+02,smith03,smith06}.  \citet{zethson99}
pointed out that the equatorial skirt is very lopsided, concentrated
on our side of the Homunculus.  If one fits an ellipse to the outer
boundary of the skirt and centers this ellipse on the star, it is
evident that the parts of the skirt on the far side of the Homunculus
are missing (see their Figure 1).  If one fits an ellipse to the
observed boundaries of the skirt, the center of such an ellipse is
more than 1$\arcsec$ NW of the star (again, see Fig.~1 in
\citealt{zethson99}).  Instead of a full azimuthally symmetric disk
(like a ``tutu'' at the waist of the Homunculus), the observable
equatorial skirt is more like a wide pie wedge; this is represented by
the blue-green area in Figure~\ref{fig:sketch}.  This makes sense in
light of the new ALMA data that show a pronounced gap in the NW side
of the inner IR/CO torus.  This gap may allow visible and near-UV
light to escape preferentially in these directions to illuminate the
skirt.

This preferred direction for all these equatorial features that point
in our direction would seem to violate the Copernican principle,
unless they are all physically related to the orientation of the orbit
of the central eccentric binary system (which, by chance, does happen
to point toward us in the equatorial plane).  As noted earlier, the $e
\simeq 0.9$ orbit ($a$=15 au) of the central binary observed at the
present epoch has the secondary star located to the NW of the primary,
so that the major axis of the orbit points out toward the NW in the
equatorial plane \citep{madura12}.

There are key ways in which the orientation of the eccentric orbit may
have played a role in shaping the ejecta.  First, at early times
during the 19th century eruption, the presence of the massive
companion star in its orbit may have actually disrupted the density
structure of the torus.  At the earliest phases after ejection of the
Homunculus in the late 1840s \citep{morse01,smith17}, the size of the
torus may have been comparable to the size of the orbit, and when at
apastron, the companion may have actually ripped through that torus
and punched a hole.  Later, in post eruption times, any post-eruption
wind would have had an easier time escaping through this
hole.\footnote{Note, however, that it seems unlikely that the
  directionality of the companion's wind (i.e., the opening of the
  colliding-wind shock cone) is the sole agent that shaped this gap,
  because the companion wind's momentum falls short by orders of
  magnitude.  In order for the low-density wind to be able to shape
  the equatorial torus, the equatorial material would need a total
  mass far below 0.1 $M_{\odot}$. Roughly 0.2 $M_{\odot}$ is the low
  end of values we estimate from CO emission below.}  In particular,
with the secondary spending most of its time at apastron, the shock
cone of the interacting winds is aimed in the same direction as the
gap in the torus.  This pre-existing gap may have funneled the fast
wind and UV radiation of the companion out in this direction.
Moreover, dust that was formed in the post-shock cooling zones of the
colliding-wind shock cone \citep{smith10} would travel downstream
preferentially in the NW direction, out through this hole.

As noted earlier, we suspect that the complex of dust clumps that
composes the Weigelt knots (and related features in the hot dust peak
NW of the star; \citealt{ches05}) may be from this type of dust
formation in the post-eruption colliding winds that piles up at this
location.  As such, the peak of hot dust immediately NW of the star
(Figure~\ref{fig:IR}c) may actually be unrelated to the rest of the
cool dust and the molecular torus.  In fact, its location would place
it inside the ring that would define the location of the torus at the
intersection of the two polar lobes (Figure~\ref{fig:sketch}).
Moreover, we know that this feature is younger than the rest of the
Homunculus anyway, as noted above \citep{dorland04,smith04a}.  In a
similar vein, elongated features like the Sr filament or other
structures in front of the NW polar lobe may be caused by ablation
from this post-eruption dust ejection that gets entrained by the
faster wind escaping in this direction.  The peculiar excitation of
the Sr filament may be due partly to the fact that it is in the shadow
of the hot dust clumps closer to the star.

The pronounced gap in the density structure on the blue side of the
torus and the associated blowout is also interesting in relation to
the structure of the polar lobes. Specifically, the polar lobes are
very nearly axisymmetric --- their geometry {\it does not} exhibit
blowouts on the side facing Earth \citep{kd01,smith06}.  Why not?  If
the torus is the thing that constricted the waist of the Homunculus to
form the bipolar structure, then presumably a much lower density in
the torus in one direction would give it much less inertia with which
to pinch the waist there.  We must conclude that either (1) the IR/CO
torus is not the agent that constricted the waist of the lobes, and
the bipolar shape has an origin in the ejection from the star itself
\citep{smith06,smith03}, or (2) the disruption of the NW portion of
the torus by the companion happened after the Homunculus polar lobes
were already shaped by it at very early times, and at radii smaller
than the apastron distance of the companion (or both).  We do not
propose a clear answer here, but this is an important consideration
for any explanation for the shaping of the nebula that invokes a
binary (or triple) system.

\subsection{Mass estimate from CO data}

As noted above, the CO emission coincides with the cooler dust in the
very dense walls of the polar lobes, where we expect the gas and dust
temperature (140~K) to be similar.  The peak CO brightness temperature
we observe is 12.4 K, implying either that the CO is very optically
thin, or if it is optically thick, that it most likely has a filling
factor less than 0.1 in our beam.  Averaged over a projected area on
the sky of 450 arcsec$^2$ (0.056 pc$^2$ at the 2.3 kpc distance of
$\eta$ Car; \citealt{smith06}), we measure an average integrated
surface brightness in $^{12}$CO 2$-$1 emission of 26 ($\pm$0.4) K km
s$^{-1}$ in our ALMA data (with a peak value of 397 K km s$^{-1}$).
Converting this to a mass of molecular gas around $\eta$~Car is
tricky.  One might estimate the total molecular hydrogen gas mass from
the integrated CO 2$-$1 emission using a standard X-factor approach,
assuming the `average' X-factor from \citet{bolatto13}, where $X_{CO}$
= 2$\times$10$^{20}$ cm$^{-2}$ (K km s$^{-1}$)$^{-1}$.  This is the
X-factor for $^{12}$CO (1-0); the (2-1) X-factor is similar, perhaps
80\% of the 1-0 value \citep{leroy13}, which would lower our value a
little.  Using this assumption, the corresponding total inferred
molecular hydrogen mass in the area of the torus would be about
5~$M_{\odot}$ or higher depending on the relative CNO abundance.
However, this value uses the standard X-factor intended for average
measurements of a typical spiral galaxy, which is probably
inapplicable here.

The high density and intense radiation field in the Homunculus are, to
say the least, very different from typical galaxy ISM conditions, so
the appropriate X-factor to use here may be different.  For example,
for the dense and warm ISM in starburst galaxies, \citet{bolatto13}
suggest using $X_{CO}$ = 0.4$\times$10$^{20}$ cm$^{-2}$ (K km
s$^{-1}$)$^{-1}$, which is a factor of 5 lower, implying a total mass
for the torus around $\eta$~Car of 1 $M_{\odot}$ (or somewhat more
depending, again, on the relative CNO abundance).  In any case, the
correct X-factor to use is highly uncertain.  An LTE optically thin
analysis leads to a total CO mass of roughly 5.9$\times$10$^{-4}$
$M_{\odot}$.  Converting this to a total molecular H gas mass assuming
a standard CO/H$_2$ number ratio of 10$^{-4}$ gives roughly
0.42~$M_{\odot}$.  On the other hand, \citet{nota02} estimated a
CO/H$_2$ mass ratio of 2.3$\times$10$^{-3}$ (number ratio of
1.6$\times$10$^{-4}$) in the dense, N-enriched, and irradiated
circumstellar environment of the LBV star AG~Carinae.  Using this
latter conversion would imply a lower total molecular gas mass in
$\eta$~Car's torus of only 0.26~$M_{\odot}$.  Clearly the dominant
uncertainty here is in the appropriate conversion of CO to H$_2$, with
implied total H$_2$ masses ranging from 0.2 to 5 $M_{\odot}$ depending
on the adopted conversion.  Consequently, we infer $\sim$1 $M_{\odot}$
as a very rough order-of-magnitude estimate for the total molecular
gas mass of the CO torus seen in our ALMA data, although we
acknowledge that further study including a multi-level analysis of CO
is needed to understand the implications of the CO emission in this
particular environment.

Detailed study of N-rich molecular chemistry in warm CSM near a
luminous star is beyond the scope of this paper, but is needed before
one can translate the CO measurements to a reliable mass.  A total
mass of as much as several $M_{\odot}$ in the equator is reasonably in
line with what one expects for the equatorial portion of the
Homunculus \citep{smith03,sf07}, which should have roughly 10-20\% of
the total mass of the Homunculus.  Although we detect some CO from the
inner side walls of the polar lobes, much deeper observations are needed to
investigate the mass in the fainter and cooler caps of the polar
lobes.

\subsection{One-sided equatorial mass ejection}

Recent studies of Type IIn supernovae (SNe) have given some examples
where the observed signatures of strong interaction with dense
circumstellar material (CSM) show evidence of a blast wave interacting
with an equatorial disk or torus.  This is based on multiple-peaked
line profiles at late times, polarization, or other evidence
\citep{leonard00,hoffman08,mauerhan14,smp14,emily14}. In some cases,
the mass distribution around the torus is thought to significantly
break azimuthal symmetry.  This presents an interesting mystery,
because equatorial mass loss from rotating stars or from close
mass-transferring binaries should be azimuthally symmetric like the
ring around SN~1987A.  Some examples with such azimuthal asymmetry are
SN~1998S \citep{leonard00,ms12,shivvers15}, SN~2010jl
\citep{fransson14}, PTF11iqb \citep{smith15}, SN~2012ab
\citep{bilinski17}, SN2013L \citep{andrews17}, and others.  Episodic
mass ejection and eccentric binaries have been invoked as possible
explanations for some of these events.

The CO torus around $\eta$~Car may provide us with a vivid pre-SN
snapshot of this type of non-azimuthally symmetric CSM distribution.
As such, it may give confirmation of the suspected influence of mass
ejection in eccentric binary systems.  Imagine that another few
hundred years pass by.  The fast polar lobes of the Homunculus will
expand to very large distances, and will become optically thin and
faint.  The slower and denser equatorial torus seen in ALMA data
presented here, however, will linger near the star (concentrated on
one side), waiting to be hit by the eventual SN blast wave.  When this
finally happens, the collision with the denser redshifted side of the
torus will have much more intense CSM interaction luminosity than the
near side, which has been mostly blown out as noted above.  As such,
the resulting emission line profile from the interaction
(i.e. H$\alpha$) will not be a symmetric double-peaked line that we
expect from a torus. Rather, the red peak will be much stronger,
resembling observed cases of SNe like PTF11iqb \citep{smith15}.
Interestingly, there is strong evidence that $\eta$~Car has suffered
mostly one-sided mass ejections in its past as well \citep{kiminki16},
although with a different orientation.

The total mass of dense molecular gas that we infer from the CO
observations (very roughly, 0.2-5 $M_{\odot}$; see above) overlaps
well with CSM mass estimates from SNe~IIn with signs of strong CSM
interaction (see \citealt{smith16} for a review).  This would be the
case if $\eta$~Car explodes in the future after a long delay, and
interacts primarily with its slow equatorial gas that remains in the
vicinity of the star.  Had it exploded a decade or so after the Great
Eruption, the gas in the Homunculus polar lobes would have been much
closer to the star and would have been hit by the SN right away,
producing a super-luminous Type II event.  The asymmetry in that case
would probably be much less pronounced, due to the high degree of
axial symmetry in the polar lobes around $\eta$~Car.

\section*{Acknowledgements}

This paper makes use of the following ALMA data:
ADS/JAO.ALMA\#2013.1.00661.S.  ALMA is a partnership of NSF (USA), ESO
(representing its member states), and NINS (Japan), together with NRC
(Canada), MOST and ASIAA (Taiwan), and KASI (Republic of Korea), in
cooperation with the Republic of Chile. The Joint ALMA Observatory is
operated by ESO, AUI/NRAO, and NAOJ.  The National Radio Astronomy
Observatory (NRAO) is a facility of the National Science Foundation
(NSF) operated under cooperative agreement by Associated Universities,
Inc.  N.S.'s research on Eta Carinae and eruptive massive stars was
supported by NSF grants AST-1312221 and AST-1515559.  Additional
support for this work was provided by NASA grants AR-12618 and
AR-14586 from the Space Telescope Science Institute, which is operated
by the Association of Universities for Research in Astronomy,
Inc. under NASA contract NAS 5-26555.

\end{document}